\documentclass[aip,amsmath,amssymb,reprint]{revtex4-1}

\usepackage{graphicx}
\usepackage{dcolumn}
\usepackage{bm}

\begin{document}


\title{Optimal analysis of ultra broadband energy-time entanglement for high bit-rate dense wavelength division multiplexed quantum networks}

\author{F. Kaiser, D. Aktas, B. Fedrici, T. Lunghi, L. Labont\'e, and S. Tanzilli}
\email{sebastien.tanzilli@unice.fr.}
\affiliation{Universit\'e Nice Sophia Antipolis, Laboratoire de Physique de la Mati\`ere Condens\'ee,\\CNRS UMR 7336, Parc Valrose, 06108 Nice Cedex 2, France}

\date{\today}

\begin{abstract}
We demonstrate an experimental method for measuring energy-time entanglement over almost $80\rm\,nm$ spectral bandwidth in a single shot with a quantum bit error rate below 0.5\%.
Our scheme is extremely cost-effective and efficient in terms of resources as it employs only one source of entangled photons and one fixed unbalanced interferometer per phase-coded analysis basis.
We show that the maximum analysis spectral bandwidth is obtained when the analysis interferometers are properly unbalanced, a strategy which can be straightforwardly applied to most of today's experiments based on energy-time and time-bin entanglement.
Our scheme has therefore a great potential for boosting bit rates and reducing the resource overhead of future entanglement-based quantum key distribution systems.
\end{abstract}

\maketitle

\section{Introduction}
To date, quantum key distribution (QKD) provides a powerful means to establish provably secure communication~\cite{Gisin02_QC,Scarani09_securityQKD,Vazirani_PRL_14}. In this perspective, QKD systems have already been commercialized, and laboratory demonstrations have achieved bit rates up to $\sim 1\rm\,Mbit/s$ at a distance of 50\,km~\cite{Lucamarini_50km_1Mbit_2013,Zhong_50km_1Mbit_2015}, extendible to up to 307\,km~\cite{Korzh_QKD300km_2015}.
Most of the reported approaches are based on laser pulses attenuated down to the single photon level.

In order to increase these rates, several multiplexing techniques can be exploited~\cite{Walborn_hyper_2008,Lucamarini_50km_1Mbit_2013,DWDM-QKD-Proposal,Ghalbouni_DWDM_OL_2013,Aktas_source_2016,Reimer_comb_2016}. Here, we focus on dense wavelength division multiplexing (DWDM), where similarly to today's classical telecommunication systems, $N$ signals at separate wavelengths can be multiplexed in, and demultiplexed out of, a fiber link, thus increasing the achievable bit rate by a factor $N$. However, for protocols based on faint laser pulses or single photons, this requires employing $N$ sources, and, depending on the protocol, up to $N$ analyzers, which strongly increases the resource overhead~\cite{Faint_Pulses_DWDM}.

In this perspective, entanglement-based DWDM QKD has the potential of increasing bit rates with significantly less technological resources.
Actually, particular \textit{single} sources of entangled photon pairs can naturally generate a broadband flux of wavelength correlated photon pairs which can be demultiplexed into $N$ correlated pairs of wavelength channels~\cite{DWDM-QKD-Proposal,Ghalbouni_DWDM_OL_2013,Aktas_source_2016,Reimer_comb_2016}.
Additionally, in comparison to QKD schemes based on laser pulses, entanglement-based approaches are compatible with device-independent network-enabling protocols~\cite{Brunner_Bell_14,SimonRepeater11,Kocsis_Heralded_Amplification,AnthoAmpli13,Simon07}. Moreover, this architecture is immune to side channel attacks, making it more robust for secure communication~\cite{Vazirani_PRL_14}.

Although long distance entanglement distribution has already been demonstrated~\cite{Takesue13_300km,Aktas_source_2016}, only a few experiments have considered DWDM QKD with, to date, in up to eight channel pairs~\citep{Aktas_source_2016}. However, in all of these realizations, the optimal analyzer settings showed a strong wavelength dependence, such that entanglement has been measured in multiple channel pairs sequentially~\cite{Lim10,Monolithic_DWDM,Aktas_source_2016}, \textit{i.e.} the analyzer settings had to be adapted for each individual channel pair. Considering that QKD implementations require measuring entanglement in two orthogonal bases, this implies that each user has to employ $2 \times N$ long-term stable individual analyzers, which is both impractical and resource demanding.

In principle, entanglement can be distributed using any observable. However, long-distance implementations often rely on energy-time or time-bin entanglement~\cite{Tittel_balanced_1999,Cuevas_balanced_2013,Antho12_CrossTB,Franson,Kwiat_EnergyTime,TimeBin_50km}, due to their immunity against polarization mode dispersion and drifts~\cite{hubel_high-fidelity_2007}.

In this paper, we demonstrate a scheme that requires only one entangled photon pair source and one analyzer per user and analysis basis to measure energy-time entanglement with less than 0.5\% error rate in a single shot over a spectral bandwidth of $\sim 80\rm\,nm$, corresponding to $N=46$ standard 100\,GHz telecommunication channel pairs.
We find that, compared to the standard configuration with identically unbalanced interferometers~\cite{Kwiat_EnergyTime,DWDM-QKD-Proposal}, the number of exploitable channel pairs can be augmented by three times when properly detuning one of the analyzers.
This represents a significant step towards cost-effective entanglement-based high bit rate QKD in DWDM networks.

\section{Experimental setup}

\begin{figure}[t!]
\centering
\includegraphics[width=8cm]{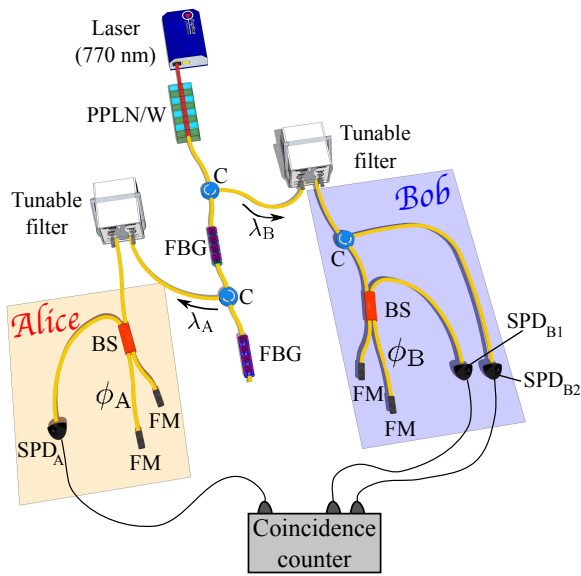}
\caption{\textbf{Experimental setup based on the Franson configuration\cite{Franson}.} The energy-time entangled photon pair source is made of a pump laser and a PPLN/W. Long (short) wavelength photons are sent to Alice (Bob) using two FBGs. Each user employs an unbalanced fiber interferometer for entanglement analysis. The interferometers' path length differences can be fine tuned using piezoelectric fiber stretchers. \label{Setup}}
\end{figure}

\begin{figure}[t!]
\centering
\includegraphics[width=7cm]{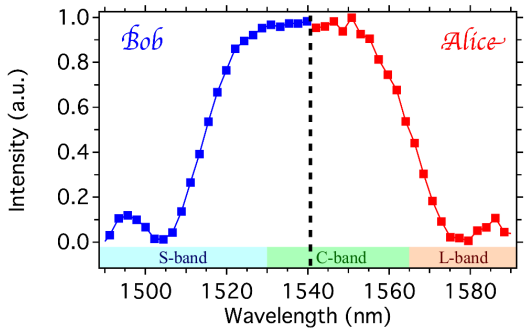}
\caption{\textbf{Emission spectrum of the PPLN/W.} The 55\,nm broad emission spectrum covers the full telecom C-band, as well as parts of the adjacent S- and L-bands. Photon pairs are generated pairwise symmetrically apart from the degenerate wavelength of 1540\,nm. Photons above and below 1540\,nm are sent to Alice and Bob, respectively. \label{Spectrum}}
\end{figure}

The experimental setup is shown in \figurename~\ref{Setup}.
A continuous-wave laser operating at $\lambda_{\rm p}=770\, \rm nm$ with a coherence length of $\sim 250\rm\,m$ pumps a periodically poled lithium niobate waveguide (PPLN/W) in which energy-time entangled photon pairs are generated around the degenerate wavelength of 1540\,nm by spontaneous parametric down-conversion (SPDC). The emitted photon pairs are directly collected into a butt-coupled single-mode fiber. The corresponding emission spectrum is shown in \figurename~\ref{Spectrum}, for which quasi phase-matching engineering leads to a bandwidth of 55\,nm $(\leftrightarrow 7\rm \, THz)$ which fully covers the commonly used telecom C-band ($1530-1565\rm\,nm)$, as well as parts of the adjacent S- ($1460-1530\rm\,nm$) and L-bands ($1565-1625\rm\,nm$).
Due to conservation of the energy during the SPDC process, the wavelengths of the paired photons $(\lambda_{\rm A,B})$ are related to the pump laser wavelength through the following relation:
\begin{equation}
\frac{1}{\lambda_{\rm p}} = \frac{1}{\lambda_{\rm A}} + \frac{1}{\lambda_{\rm B}}.\label{Energy}
\end{equation}
In other words, the photons are generated pairwise symmetrically apart from the degenerate wavelength.
The pairs are deterministically separated by sending long (short) wavelength photons to Alice (Bob) using a set of two broadband fiber Bragg gratings (FBG) and associated circulators (C).
Further dynamic wavelength filtering is achieved using two tunable filters with a 0.8\,nm ($\leftrightarrow 100\rm\,GHz$) flat-top transmission profile, mimicking standard 100\,GHz dense wavelength division multiplexers.
To reveal energy-time entanglement, Alice and Bob use each an unbalanced fiber interferometer (Franson configuration~\cite{Franson}), made of a beam-splitter and two Faraday mirrors. Both interferometers have a path length difference of $\Delta L_{\rm A,B} \approx 6.7\rm\,cm$, can be fine tuned, and are actively stabilized using piezoelectric fiber stretchers~\cite{TheKaiser14Long}.
At the interferometer output, Alice detects her photons using a free-running InGaAs single photon detector (SPD, IDQ id220, 20\% detection efficiency). At Bob's site, we use an additional circulator through which we can detect photons at both interferometer outputs using gated InGaAs SPDs (IDQ id201, 25\% detection efficiency).\\
Concerning the quantum state of the photon pairs at the interferometers' outputs, four contributions have to be considered. Either both photons take the short or long paths ($s_{\rm A}-s_{\rm B}$ or $l_{\rm A}-l_{\rm B}$), or both photons take opposite paths ($s_{\rm A}-l_{\rm B}$ or $l_{\rm A}-s_{\rm B}$).
Due to the spontaneous character of the photon pair generation process, the pair creation time in the PPLN/W remains unknown. This makes the contributions $s_{\rm A}-s_{\rm B}$ and $l_{\rm A}-l_{\rm B}$ indistinguishable, which leads to the observation of entanglement~\cite{Kwiat_EnergyTime}.
These contributions are selected using a fast coincidence logic, leading to a reduced quantum state
\begin{equation}
|\psi \rangle_{\rm post} = \frac{1}{\sqrt{2}} \left( | s_{\rm A} \rangle | s_{\rm B} \rangle + {\rm e}^{{\rm i} \, \phi} | l_{\rm A} \rangle | l_{\rm B} \rangle\right),
\end{equation}
where $\phi = \phi_{\rm A} + \phi_{\rm B}$ stands as the two-photon phase. The individual contributions, $\phi_{\rm A,B}$, are related to the interferometers' path length differences by
\begin{equation}
\phi_{\rm A,B} = \frac{2\pi\,\Delta L_{\rm A,B} \, n(\lambda_{\rm A,B})}{\lambda_{\rm A,B}}.\label{Phases}
\end{equation}
Here, $n(\lambda_{\rm A,B})$ is the wavelength-dependent refractive index of the fibers in the interferometers.
According to reference~\cite{Kwiat_EnergyTime}, the rate of coincidences between detectors $\rm SPD_A$ and $\rm SPD_{B_1}$ follows $R_{\rm AB_1} \propto 1 + \cos \phi$, while the rate between detectors $\rm SPD_A$ and $\rm SPD_{B_2}$ follows $R_{\rm AB_2} \propto 1 - \cos \phi$.
For entanglement-based QKD using the Ekert protocol~\cite{Ekert91}, the analysis bases are defined by the following settings:
\begin{eqnarray}
\phi_{\rm A} + \phi_{\rm B} &=& 0\label{ZeroPhase}\\
 \phi'_{\rm A} + \phi'_{\rm B} &=& \pi \qquad {\rm with} \qquad \phi'_{\rm A} = \phi_{\rm A} + \frac{\pi}{2}.\label{SecondBasis}
\end{eqnarray}
In general, these conditions cannot be fulfilled over a large spectral bandwidth for fixed path length differences ($\Delta L_{\rm A,B}$), which results in a wavelength dependent two-photon phase shift. Considering the setting given in equation~\ref{ZeroPhase}, this leads to an undesired non-zero anti-coincidence rate $R_{\rm AB_2}$. The associated quantum bit error rate (QBER) of the QKD link is then given by
\begin{equation}
{\rm QBER} = \frac{R_{\rm AB_2}}{R_{\rm AB_1}+R_{\rm AB_2}} = \sin^2 \left( \frac{\phi}{2}\right).\label{QBER}
\end{equation}
Although there exist several algorithms for QBER correction~\cite{Gisin02_QC,Scarani09_securityQKD}, they usually require additional resources, having repercussions on the attainable bit rate of the QKD link. Therefore, it is commonly acknowledged to keep the QBER as low as possible~\cite{Gisin02_QC,Scarani09_securityQKD,Vazirani_PRL_14}. Additionally, it has been demonstrated that high-dimensional QKD is only efficient at a QBER very close to zero~\cite{Lucamarini_50km_1Mbit_2013}. 
For this reason, we fix the maximum allowed QBER induced by improper interferometer settings to a stringent value of 0.5\%, corresponding to an acceptable two-photon phase shift of $\phi = \pm 0.14\,\rm rad$.\\
\,\\
\textbf{Spectral dependence of the two-photon phase $\phi$}

To calculate the spectral dependence of $\phi$, we first express $n(\lambda)$ by a second order Taylor series, which reads
\begin{equation}
n(\lambda) \approx n_0 + \frac{dn}{d \lambda} \cdot \Delta \lambda + \frac{1}{2} \frac{d^2n}{d \lambda^2} \cdot (\Delta \lambda)^2.\label{RefIndex}
\end{equation}
Here, $n_0$, $\frac{dn}{d \lambda}$, and $\frac{d^2n}{d \lambda^2}$ are the fiber refractive index, the first, and second order derivatives, respectively, at the degenerate wavelength ($2\lambda_{\rm p}$). All coefficients can be inferred from Sellmeier equations~\cite{leviton_temperature-dependent_2008}.
By inserting equation~\ref{RefIndex} into equation~\ref{Phases}, and respecting equation~\ref{Energy}, we obtain
\begin{equation}
\phi = \phi \left(n_0,\frac{dn}{d \lambda},\frac{d^2n}{d \lambda^2},\Delta L_{\rm A},\Delta L_{\rm B},\lambda_{\rm A},\lambda_{\rm p} \right).\label{TotalPhase}
\end{equation}
It is often considered that the QBER is minimized for $\Delta L_{\rm A} = \Delta L_{\rm B}$~\cite{Tittel_balanced_1999,Cuevas_balanced_2013}. However, in section \ref{Non_identical_theory} we show that the optimal settings are rather obtained for identical path travel time differences.
In this case of $\Delta L_{\rm A} = \Delta L_{\rm B}$, equation~\ref{TotalPhase} simplifies to
\begin{equation}
\phi = \frac{d^2n}{d \lambda^2} \cdot \frac{\pi (\lambda_{\rm A} - 2\lambda_{\rm p})^2}{\lambda_{\rm A} - \lambda_{\rm p}} \cdot \Delta L_{\rm A} + \mathcal{C},\label{BalancedPhase}
\end{equation}
in which $\mathcal{C} = \frac{2 \pi \cdot n_0 \cdot \Delta L_{\rm A}}{\lambda_{\rm p}}$ is a constant as it is independent of the wavelengths of the paired photons ($\lambda_{\rm A,B}$).

\section{Results with identical analyzers}

\begin{figure}[t!]
\centering
\includegraphics[width=7cm]{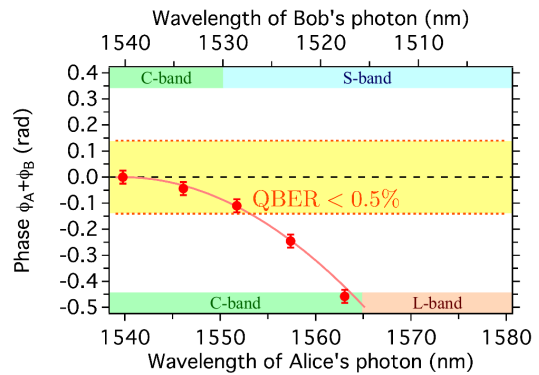}
\caption{\textbf{Two-photon phase shift for identical analysis interferometers.} The yellow shaded area indicates the region in which the QBER stays below 0.5\%.
The exploitable bandwidth covers 16 pairs of standard 100\,GHz telecommunication channels.
Error bars assume poissonian photon number statistics. The fit to the data is obtained with equation~\ref{BalancedPhase}.\label{ResultsEqual}}
\end{figure}

For different wavelengths $\lambda_{\rm A}$ (and symmetrically associated wavelengths $\lambda_{\rm B}$), we infer the two-photon phase $\phi$ by measuring the QBER and solving equation~\ref{QBER}.
We first align both interferometers to exactly $\Delta L_{\rm A} = \Delta L_{\rm B}$. This is done by using an iterative procedure that amounts to infer the wavelength dependence of the QBER for different $\Delta L_{\rm A}$ until a flat distribution is found around 1540\,nm. Experimental results for $\Delta L_{\rm A} = \Delta L_{\rm B}$ are shown in \figurename~\ref{ResultsEqual}. Starting with $\phi=0$ at $\lambda_{\rm A} = 1540\rm\,nm$, we reach the threshold phase shift $\phi=-0.14\,\rm rad$ at $\lambda_{\rm A} \sim 1553\rm\,nm$ ($\lambda_{\rm B} \sim 1527\rm\,nm$).
Consequently, these interferometers can be used to analyse entanglement with a $\rm QBER<0.5\%$ for $1540\,{\rm nm} < \lambda_{\rm A} < 1553\,{\rm nm}$ ($1527\,{\rm nm} < \lambda_{\rm B} < 1540\,{\rm nm}$) simultaneously, which corresponds to 16 pairs of standard 100\,GHz telecommunication channels~\cite{ITU,Aktas_source_2016}.
Note that fully exploiting the emission bandwidth of our photon pair source requires analysing entanglement in at least 40 pairs of 100\,GHz channels simultaneously.
A straightforward solution would be to employ analysis interferometers made of custom-made components to shift or compensate dispersion. For example, by employing hybrid interferometers, made of single-mode and dispersion compensation fibers, instead of fully single-mode fiber interferometers, a 1.4\% increase in the interference visibility was observed over a spectral bandwidth of 1.6\,nm~\cite{Zhong_nonlocal_cancellation_2013}.

\section{Results with optimal analyzers \label{Non_identical_theory}}

In the following, we demonstrate a much cheaper and simpler approach which can be applied without changing any of the components in the standard setup. We note that our strategy is not limited to fiber-based analysis interferometers only. For example, planar lightwave circuits~\cite{Korzh_QKD300km_2015}, where dispersion compensation is not straightforward, could also benefit from the proposed method.
Let us consider the different central wavelengths of Alice's and Bob's photons, $\lambda^*_{\rm A} \sim 1560\,\rm nm$ and $\lambda^*_{\rm B} \sim 1521\,\rm nm$. Note that these spectral contributions show different group velocities
\begin{equation}
v_{\rm A,B} = \frac{c}{n_0-\dfrac{d n}{d \lambda}\cdot \lambda^*_{A,B}}.
\end{equation}
As a consequence, the wavepackets in Alice's and Bob's interferometers show non-identical path travel time differences between short and long arms.
To match these time differences, the following equation needs to be fulfilled
\begin{equation}
\dfrac{\Delta L_{\rm A}}{v_{\rm A}} = \dfrac{\Delta L_{\rm B}}{v_{\rm B}}. \label{GroupMatchedInterferometers}
\end{equation}
Using Sellmeier equations~\cite{leviton_temperature-dependent_2008} to infer $n_0$ and $\frac{dn}{d \lambda}$, and $\Delta L_{\rm B} = 6.7\rm\,cm$, we fulfill equation~\ref{GroupMatchedInterferometers} by setting the path length difference of Alice's interferometer to $\Delta L_{\rm A} = \left(\Delta L_{\rm B} -  12\rm\,\mu m \right)$.
\begin{figure}[t!]
\centering
\includegraphics[width=7cm]{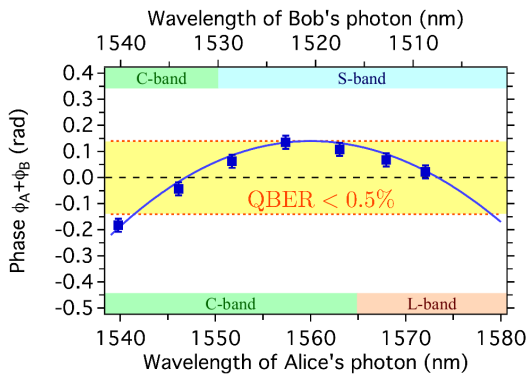}
\caption{\textbf{Two-photon phase shift for non-identical analysis interferometers.} The path length difference of Alice's interferometer has been reduced by $\sim 12\rm\,\mu m$ compared to Bob's. Now, the region in which the QBER is below 0.5\% covers 46 pairs of standard telecommunication channels, meaning that the full telecom C-band, as well as some parts of the adjacent S- and L-bands, can be exploited simultaneously for entanglement analysis.\label{ResultsUnequal}}
\end{figure}
The associated experimental results are shown in \figurename~\ref{ResultsUnequal}. We have essentially shifted the curve in \figurename~\ref{ResultsEqual} by about $20\,\rm nm$ to $\lambda^*_{\rm A} \sim 1560\,\rm nm$ ($\lambda^*_{\rm B} \sim 1521\,\rm nm$). This way, we are now able to keep the two-photon phase within $\phi = \pm 0.14\,\rm rad$ for the full emission bandwidth of our source ($1541\,{\rm nm} < \lambda_{\rm A} < 1579\,{\rm nm}$ and $1503\,{\rm nm} < \lambda_{\rm B} < 1539\,{\rm nm}$), allowing to analyze entanglement with a QBER below 0.5\% in 46 pairs of standard 100\,GHz telecommunication channels in a single shot. We note that the improved bandwidth fully covers the most commonly used telecom C-band, as well as some parts of the adjacent S- and L-bands.

We envision the following configuration for realizing future DWDM QKD links. The tunable bandpass filters will be removed and Alice is also supplied with an interferometer having two outputs (as Bob's in our current configuration, see \figurename~\ref{Setup}). Wavelength division multiplexing is performed after the interferometers using standard telecom $N$-channel DWDMs.
Granting security of DWDM QKD protocols requires using a second analysis basis (see equation~\ref{SecondBasis}). This can be implemented by providing each Alice and Bob with a second interferometer for which the path length differences are set to $\Delta L'_{\rm A,B} = \Delta L_{\rm A,B} + \frac{\lambda_{\rm p}}{2\cdot n_0}$. In this case, group velocity dispersion causes a slight additional error in $\phi'$ over the full bandwidth of the source. From equation~\ref{Phases} we calculate that it will be below $\frac{2\pi}{300}$, which is essentially negligible for choosing two complementary bases.
In a long-distance scenario, group velocity dispersion in a standard fiber distribution link causes a broadening of the coincidence peaks such that the contributions $s_{\rm A} - s_{\rm B}$ and $l_{\rm A} - l_{\rm B}$ cannot be properly post-selected. However, this problem can be conveniently overcome using dispersion compensation~\cite{Fasel_30km_2004,Aktas_source_2016} and/or dispersion shifted fibers~\cite{Marcikic_Tele_2003}.

Finally, we stress that our approach can be also applied to optimize DWDM QKD with polarization entangled photon pairs where wavelength dependent birefringence in the half-wave plates is an issue~\cite{Lim10}. Analogously to the strategy for energy-time entangled photon pairs, this problem could be partially compensated by either using thicker/thinner half-wave plates (optimized for longer/shorter wavelengths), or by simply tilting the existing ones.

\section{Conclusion}

In conclusion, we analysed energy-time entanglement of a broadband photon pair source using fixed unbalanced fiber interferometers, in the perspective of DWDM QKD.
In the standard configuration, with identical analysis interferometers at Alice's and Bob's sites, group velocity dispersion limits the analysis bandwidth to 16 standard 100\,GHz channel pairs at a QBER threshold of 0.5\%.

Without replacing any components in the experimental setup, solely by properly unbalancing one of the two interferometers, we improved the analysis bandwidth to 46 channel pairs, covering not only the commonly used telecom C-band, but also some of the adjacent S- and L-bands.

We stress that the number of channel pairs could be increased to 368 by using the 12.5\,GHz ultra-dense DWDM grid, which underlines the tremendous potential for high bit rate entanglement-based DWDM QKD.

Therefore, we believe that our work will have a great impact for the optimal exploitation of current and future high bit rate DWDM QKD systems, especially in combination with other multiplexing techniques.

\section{Acknowledgments}

The authors acknowledge financial support from the Foundation Simone \& Cino Del Duca, the European Commission for the FP7-ITN PICQUE project (grant agreement No 608062), l'Agence Nationale de la Recherche (ANR) for the CONNEQT and SPOCQ projects (grants ANR-EMMA-002-01 and ANR-14- CE32-0019, respectively), and the iXCore Research Foundation.

%


\end{document}